\documentclass [11pt]{article}
\usepackage{amsmath,amsthm,amsfonts,amscd,eucal,latexsym,amssymb,mathrsfs,cancel}
\usepackage{epsfig}  
\usepackage{simplewick,bm}
\usepackage{hyperref}
\usepackage{color}
\input xy
\xyoption{all}

\def\beq{\begin{eqnarray}}
\def\eeq{\end{eqnarray}}

\def\WF{{\text{WF}}}

\newtheorem*{theorem*}{Theorem}

\theoremstyle{definition}
\newtheorem{definition}{Definition}[section]

\begin{document}

%
%

\par
\bigskip
\Large
\noindent
{\bf Local incompatibility of the microlocal spectrum condition with the KMS property along spacelike directions in quantum field theory on curved spacetime}
\bigskip
\par
\rm
\normalsize


\large
\noindent
{\bf Nicola Pinamonti$^{1,2,a}$}, {\bf Ko Sanders$^{3,b}$}, {\bf Rainer Verch$^{4,c}$} \\
\par
\small
\noindent$^1$ Dipartimento di Matematica, Universit\`a di Genova - Via Dodecaneso, 35, I-16146 Genova, Italy. \smallskip

\noindent$^2$ Istituto Nazionale di Fisica Nucleare - Sezione di Genova, Via Dodecaneso, 33 I-16146 Genova, Italy. \smallskip

\noindent$^3$ School of Mathematical Sciences and Centre for Astrophysics and Relativity, Dublin City University, Glasnevin, Dublin 9,
Ireland.\smallskip

\noindent$^4$ Institute for Theoretical Physics, University of Leipzig, D-04009 Leipzig, Germany.\smallskip

\smallskip

\noindent E-mail:
$^a$pinamont@dima.unige.it,
$^b$jacobus.sanders@dcu.ie,
$^c$rainer.verch@uni-leipzig.de\\

\normalsize

\par

\rm\normalsize

\rm\normalsize


\par
\bigskip

\rm\normalsize
\noindent {\small Version of \today}

\par
\bigskip

\rm\normalsize

\bigskip

\noindent
\small
{\bf Abstract}
States of a generic quantum field theory on a curved spacetime  are considered which satisfy
the KMS condition with respect to an evolution associated with a complete (Killing) vector field.
It is shown that at any point where the vector field is spacelike, such states cannot satisfy a
certain microlocal condition which is weaker than the microlocal spectrum condition in the case
of asymptotically free fields.
\normalsize

\vskip .3cm

\section{Introduction}
\label{intro}

In the standard framework of quantum statistical mechanics \cite{BR} the observables of a quantum system are described by the self-adjoint elements of a non-commutative
$C^*$-algebra $\mathcal{A}$ with a unit element ${\bf 1}$.
A time-evolution is described by a one-parameter group 
of $*$-automorphisms $\{\alpha_t\}_{t\in\mathbb{R}}$ acting on $\mathcal{A}$ that is (weakly) continuous in $t$.
Together, $\mathcal{A}$ and $\{\alpha_t\}_{t\in\mathbb{R}}$ form a $C^*$-dynamical system. In this framework, 
a state $\omega$ on $\mathcal{A}$ is described by a linear functional which is normalized, $\omega({\bf 1}) = 1$,
and positive, $\omega(A^*A) \ge 0$ $(A \in \mathcal{A})$. The evaluation of $\omega$ on an observable $A\in\mathcal{A}$,
denoted $\omega(A) \equiv \langle A \rangle_\omega$,
is interpreted as the expectation value. 

\bigskip

The class of all states on a $C^*$-dynamical system is very large and it is known from examples that there are states
that correspond to unphysical scenarios like infinite energy densities or infinite pressures.
For this reason it is essential to select a class of simple states that do correspond to physically realistic configuations of a system.
A particularly important class is furnished by states which describe systems in equilibrium with a thermal bath at a fixed temperature. 
Such states are characterized by a very small number of parameters, like their temperature and chemical potential. 
Thus, for such states, the expectation values of physically relevant observables are functions of these parameters, 
and furthermore, relations between different expectation values are described by equations of thermodynamical (thermostatic) nature.
Prime examples of equilibrium states are the well-known Gibbs states \cite{Haag}, which 
are used to model the canonical ensemble of quantum particles at fixed temperature in a box. 
A generalization of Gibbs states is the class of states that satisfy the  Kubo-Martin-Schwinger (KMS) condition \cite{KMS}
with respect to the system's time-evolution $\{\alpha_t\}_{t\in\mathbb{R}}$ (see Definition \ref{def:KMS}).
States which satisfy the KMS condition are called KMS states, and they can also be defined for more general settings, where e.g.\ $\mathcal{A}$ may be a more general $^*$-algebra than a $C^*$-algebra.

\bigskip

The KMS condition is so versatile that it can be used to characterize equilibrium states for non-relativistic or relativistic quantum field theories both on 
flat and curved spacetimes.
In this letter we are interested in analyzing KMS states for relativistic quantum field theories on a 
(possibly curved) spacetime $M$ with spacetime metric $g$. In this case, the simplest concept of time-evolution, generalizing the time-translations on Minkowski spacetime, is tied to spacetimes with a smooth Killing vector field $\chi$, which generates a one-parametric group of isometries for the spacetime metric. 
In order to interpret the group of isometries as time-translations, it is natural to assume that 
$\chi$ is everywhere timelike, so the spacetime is stationary.
However, it is interesting to notice that it is not uncommon for
spacetimes to admit isometry groups generated by Killing vector
fields which are timelike in some regions of spacetime and spacelike in other regions. 
The most prominent example is furnished by the rotating Kerr black hole \cite{WaldGR}, whose 
outer horizon is the bifurcate Killing horizon of a Killing vector field that is timelike 
just outside of the outer horizon, but which becomes spacelike far away from and inside of the black hole region. 
This prompts the obvious question if global KMS states with respect to the corresponding one-parameter evolution exist in such cases.

\bigskip

As will be outlined in Section \ref{fermifields} a well-defined quasifree KMS state with respect to spacelike translations can be constructed for free Fermi 
fields\footnote{Throughout this paper we will call a theory free when the dynamical fields satisfy a linear field equation.} on flat spacetime, however,
we will see that the obtained state is not admissible in a sense to be discussed below (see Section \ref{Sec:Result} for the full definition).\footnote{A very general statement on the inconsistency of the KMS condition for spacelike translations is made in \cite{B}, but there appears to be an error in the proof of Lemma 3.6 of that reference concerning entire holomorphic functions of exponential type. The statement in \cite{B} would rule out the aforementioned Fermi field KMS states for spacelike translations on the even part of the Fermi field algebra; however the result in Sec.\ \ref{fermifields} shows they exist.}
In particular, the state will not be of Hadamard type.
For free bosonic fields on Minkowski spacetime the very same construction does not lead to the definition of a well posed two-point function of a state because of infrared divergences.
For general quantum field theories on Minkowski spacetime, satisfying locality and translation invariance,  it follows from the work of Trebels 
\cite{Trebels} that a vacuum state cannot be a KMS state for the flow of a Killing field $\chi$ in regions where $\chi$ is spacelike. (This can happen when $\chi$ is timelike, e.g.\ in the Unruh effect, cf.\ \cite{WaldQFTbook}.)

In the setting of curved spacetimes, for the particular case of the Kerr spacetime and a free quantized scalar field, satisfying the covariant Klein-Gordon equation, Kay and Wald invoked a superradiance property to argue that Killing field invariant Hadamard states do not exist \cite{KW}. For a class of states which satisfy the KMS condition they note that the superradiance assumption may be dropped. Indeed, implicitly they have shown that for a quantized Klein-Gordon field on any globally hyperbolic spacetime with a Killing field $\chi$ that becomes spacelike there can be no quasifree KMS states satisfying additional regularity conditions (see the corollary to Lemma 6.2 and the first two paragraphs of Section 6.4 in loc.cit.).\footnote{We are grateful to an anonymous referee for pointing out this implicit result.}

\bigskip

In this regard, it is worth noting that the microlocal spectrum condition, a generalization of the Hadamard condition, is by now recognized to be  an unavoidable requirement for states of quantum fields to be viewed as physically realistic\cite{BFK,FV13,FV03,WaldQFTbook}. Moreover, KMS states for linear quantum fields on a stationary spacetime, where the corresponding Killing vector field is everywhere timelike, have been shown to fulfill the microlocal spectrum condition \cite{SV}. The existence of KMS states for quantized linear fields on stationary spacetimes has also been established, see \cite{Sa3} and references cited therein.

\bigskip

In this note we show that KMS states cannot be admissible in the neighborhood of points where the vector field $\chi$ is spacelike.
No particular form of field equation needs to be assumed for the quantum field, i.e.\ the argument is model-independent. The term ``admissible'', defined in Sec.\ 2, is closely related to the microlocal spectrum condition and for free fields it is equivalent to it. Neither the regularity conditions on the KMS state used in \cite{KW}, nor the global hyperbolicity of $M$ nor the Killing property of $\chi$ are required in our proof.

\bigskip

To motivate the admissibility condition that we shall impose on states of a quantum field on a curved spacetime in Sec.\ 2,
let us first briefly recall here the microlocal spectrum condition in its simplest form, i.e.\ for a scalar quantum field $\phi$.
The spacetime $M$ is assumed to be time-oriented and globally hyperbolic.
The $*$-algebra $\mathcal{A}$ of observables is
generated by the set $\{\phi(f)\ |\ f\in\mathcal{D}(M)\}$ formed by quantum field operators smeared with smooth test functions.
A state $\omega$ on $\mathcal{A}$ is determined by the  $n$-point functions
$w^\omega_n(f_1,\ldots,f_n)=\omega(\phi(f_1)\cdots\phi(f_n))$  $(f_j \in \mathcal{D}(M))$ which, by definition, are (or rather, extend to) 
distributions in $\mathcal{D}(M^n)$. The microlocal spectrum condition for a state $\omega$ is a condition of the form
$\WF(w^\omega_n)\subset \Gamma_n$ on the wave front sets of the distributions $w^\omega_n$.
We refer to \cite{Hormander} for the concept of wave front set for distributions on a manifold, and to
\cite{BFK} for the definition of the sets $\Gamma_n$ in the microlocal spectrum condition. Informally, this condition imposes
an upper bound on the allowed singularities of the $n$-point functions.

The microlocal spectrum condition is motivated by the work of Radzikowski \cite{Ra}, who showed that for a free scalar
field
the Hadamard condition \cite{KW,WaldQFTbook} on the two-point function $w^\omega_2$ is equivalent to the
microlocal condition\footnote{ Here, $\boldsymbol{0}$ denotes the zero section in $T^*M \times T^*M$.
Writing $(x,k)$ for an element in $T^*M$ means that $k \in T_x^*M$, i.e.\
$x$ denotes the point in the base manifold $M$ at which the cotangent vector $k$ is ``affixed''.}
\begin{align} \label{eqn:WF2}
\WF(w^\omega_2) =& \left\{ (x,-k;x',k')\in T^*M\times T^*M \setminus \{\boldsymbol{0}\} \ |\ (x',k') \in \mathcal{N}^-\,, \right.\notag\\
&\left. (x,k)\sim (x',k') \right\} \,,
\end{align}
where $\mathcal{N}^-$ is the set of past-directed lightlike co-vectors of $M$, i.e.\ $g^{-1}(k')$ is a lightlike vector
and $k(v) < 0$ for any future pointing tangent vector $v$ of $M$ at $x$.\footnote{Using index notation,
$g^{-1}(k)$ reads $g^{\mu \nu}k_\nu$, and $k(v)$ reads $k_\mu v^\mu$.} Furthermore,
$(x,k)\sim (x',k')$ means that $x$ and $x'$ are joined by a null geodesic $\gamma$ and
$g^{-1}(k)$ and $g^{-1}(k')$ are tangent to $\gamma$ and coincide up to parallel transport along $\gamma$. 
Analogous statements have been proved for other types of quantized fields that are subject to linear hyperbolic field
equations \cite{SV2}. Moreover, for such fields, \eqref{eqn:WF2} is equivalent to the 2-point part of the microlocal spectrum condition, $\WF(w^\omega_2)\subset \Gamma_2$ \cite{SVW}, and for any fields satisfying canonical commutation or anti-commutation relations, it is equivalent to the full microlocal spectrum condition \cite{Sa}.
The microlocal spectrum condition has also been generalized to encompass the extended algebra of
Wick polynomials of free fields on flat and curved spacetimes \cite{BFK} and the use of microlocal techniques has
opened up the possibility of a local covariant perturbative construction of interacting field theories on curved spacetimes
\cite{BF,HW1,HW2}. In \cite{FV13} it has been shown that the microlocal spectrum condition is not only sufficient, but
also necessary for a state to be well-defined on the extended algebra of Wick polynomials. In the light of these results,
demanding that states satisfy the microlocal spectrum condition can surely be viewed as non-negotiable in local covariant
quantum field theory.

Worth noting is the equality in \eqref{eqn:WF2}: the specification of $\WF(w^\omega_2)$ is a restriction both from
``above'' and ``below'', which is stronger than the condition $\WF(w^\omega_2)\subset \Gamma_2$. The admissibility
condition that we shall impose in Sec.\ 2 is in fact a lower bound on $\WF(w^\omega_2)$, namely that it contains all
lightlike co-vector pairs $(x,-k;x,k)$ with $(x,k) \in \mathcal{N}^-$. This condition follows from \eqref{eqn:WF2}, so it
holds for free fields. More generally it is a local and covariant remnant of the Lorentz covariance of the spectrum
condition familiar from quantum field theory in Minkowski spacetime.

\bigskip

In the following section we will delineate the precise assumptions and present the proof.

\section{Result} \label{Sec:Result}

We consider a spacetime $(M,g)$, where $M$ is a four dimensional, oriented and time-oriented, smooth manifold and $g$ is a Lorentzian metric of signature $(-,+,+,+)$. 
An additional assumption could be that $(M,g)$ is globally hyperbolic, but our proof below does not require it.
Furthermore, we consider a non-commutative $*$-algebra $\mathcal{A}$ generated by a unit element ${\bf 1}$ 
and ``smeared quantum field operators'' $\phi(F)$ where $F$ stands for any smooth, compactly supported section 
of a given complex vector bundle $E$ over $M$ of some finite dimension $N$.\footnote{ The $\phi(F)$ need not be 
represented as operators on some Hilbert space, but may well be elements of an ``abstract'' algebra.}
We denote by $\Gamma$ the action of a fibrewise anti-linear involution on $E$ and 
we also assume that it induces a continuous map $\Gamma: \mathcal{D}(E) \to \mathcal{D}(E)$, where $\mathcal{D}(E)$ denotes
the space of smooth, compactly supported sections of $E$, equipped in the usual manner with the $\mathcal{D}$-topology. For
simplicity we also impose $\phi(F)^* = \phi(\Gamma F)$.

\bigskip

We also assume that $E$ admits a complete smooth vector field $V$ which projects down to a smooth, complete vector field $\chi$ on $(M,g)$. 
These vector fields generate one-parameter groups of diffeomorphisms, $\{\eta_t\}_{t\in\mathbb{R}}$ on $E$ and $\{\tau_t\}_{t\in\mathbb{R}}$ on $M$.
Furthermore, we assume that $\Gamma$ commutes with $\eta_t$ and that the diffeomorphism groups induce
a one-parametric evolution group of $*$-automorphisms $\{\alpha_t\}_{t\in\mathbb{R}}$ on $\mathcal{A}$ by
\[
\alpha_t(\phi(F)) :=  \phi(\Psi_t(F)),\qquad
\Psi_t(F):=\eta_{-t}\circ F\circ \tau_t.
\]
Then $\mathcal{A}$ together with $\{\alpha_t\}_{t\in\mathbb{R}}$ is a $*$-algebraic dynamical system.
For our discussion it is not necessary to be more specific than this. In particular, we 
don't assume any field equations to be fulfilled by the $\phi(F)$.
The hypotheses we are imposing are completely general and are satisfied by various free or interacting fermionic or bosonic fields.

\bigskip

Consider a state $\omega$ on $\mathcal{A}$, i.e.\ a linear functional on 
$\mathcal{A}\to \mathbb{C}$ which is positive ($\omega(A^*A)\ge 0$ for all $A \in \mathcal{A}$) and normalized $(\omega({\bf 1})=1)$. 
Let us write the associated {\it two-point function} as
\[
w_2^\omega(F,F')  := \omega(\phi(F)\phi(F')).
\]
We assume that the map
\[
(F,F')\mapsto w_2^{\omega}(F,F')
\]
gives rise to a distribution over compactly supported smooth sections\footnote{Here we really mean $\mathcal{D}(E\boxtimes E)$, which is the completion of $\mathcal{D}(E)\otimes\mathcal{D}(E)$ in the (unique) locally convex topology derived from the (nuclear) test-function topology on $\mathcal{D}(E)$.} $\mathcal{D}(E)\otimes\mathcal{D}(E)$ of the outer product bundle $E\boxtimes E$ over $M\times M$ (see e.g.\ \cite{SV} Sec.3.3 for a definition).

\bigskip

Below, it will turn out useful to consider a Hilbert space-valued distribution associated with the two-point function $w_2^\omega$ which
is obtained by the GNS construction (see e.g.\ \cite{FH} for a description of the GNS representation of a unital 
$*$-algebra associated with a state). For the sake of self-containedness, we describe how that Hilbert space-valued distribution
is obtained. The positivity of the state implies that
\[
w_2^{\omega}(\Gamma F,F) \geq 0,
\]
so that $w_2^{\omega}$ defines a semi-definite sesquilinear form on $\mathcal{D}(E)$ by $(F,F')\mapsto w_2^{\omega}(\Gamma F,F')$. 
The sections $F\in\mathcal{D}(E)$ with $w_2^{\omega}(\Gamma F,F)=0$ form a linear space $K$, by the Cauchy-Schwarz inequality. We denote the
equivalence classes in $\mathcal{D}(E)/K$ by $[F]$ and we complete this quotient space to a Hilbert space $\mathcal{H}$ using the inner product
\[
\langle [F], [F']\rangle := w_2^{\omega}(\Gamma F,F').
\]
It will be convenient to consider the $\mathcal{H}$-valued map $\mathcal{D}(E)\to\mathcal{H}$ defined by $F\mapsto [F]$, which we will write as
\[
F\mapsto \boldsymbol{\phi}(F)\Omega.
\]
For good reason, this is reminiscent of the GNS representation; $\boldsymbol{\phi}(F)$ is the representer of $\phi(F)$ in the GNS representation
and $\Omega$ is the GNS vector.
We note that
\[
\langle \boldsymbol{\phi}(F)\Omega, \boldsymbol{\phi}(F')\Omega\rangle = w_2^{\omega}(\Gamma F,F')
\]
and that the map $F\mapsto \boldsymbol{\phi}(F)\Omega$ is an $\mathcal{H}$-valued distribution, because $w_2^{\omega}$ is a distribution; see \cite{SVW} for details.

\begin{definition}\label{def:KMS}
A state $\omega$ on $\mathcal{A}$ satisfies the {\it KMS condition} with respect to $\{\alpha_t\}_{t \in \mathbb{R}}$ at inverse temperature $\beta$ if, for every $A,B\in \mathcal{A}$ the function
\[
t\mapsto\omega(A\alpha_t(B)) \quad \quad (t \in \mathbb{R})
\]
has a continuous and bounded extension to the closed strip
\[
S_\beta = \{{\sf t} = t + is \in \mathbb{C}\,|\, 0 \le s \le \beta\,, \ t \in \mathbb{R}\}
\]
which is analytic in the open interior of $S_\beta$ and satisfies
\[
\omega(A\alpha_{t+i\beta}(B))=\omega(\alpha_t(B)A) \quad \quad (t \in \mathbb{R})\,.
\]
The state $\omega$ is then called a {\it KMS state} at inverse temperature $\beta$ with respect to the one-parametric evolution 
 group $\{\alpha_t\}_{t \in \mathbb{R}}$. (Cf.\ e.g.\ \cite{Ped} and references cited there; the KMS condition carries over from the 
 $C^*$-algebra setting in an obvious way to states and continuous automorphism groups on more general unital $*$-algebras.)
\end{definition}

One can show that KMS states are necessarily invariant under all the $\alpha_t$. This invariance is typically incompatible 
with the dynamics of the field, unless the evolution $\tau_t$ is given by isometries of the spacetime, i.e.\ $\chi$ is a Killing field. 
For our result, however, this Killing property will not be required.

If $\omega$ is a KMS state at inverse temperature $\beta$, then the 
2-point function satisfies properties analogous to those listed in Def.\ \ref{def:KMS}, 
in particular $w_2^\omega$ is invariant under the action of $\eta_t$ and $\tau_t$ in the sense that
\[
w_2^\omega(\Psi_t(F), \Psi_t(F')) =  w_2^\omega(F, F')
\]
and the KMS condition implies that for any $F,F'\in\mathcal{D}(E)$ the map
\[
t \mapsto w_2^\omega(F, \Psi_t(F'))
\]
has a bounded and continuous extension to the strip $S_\beta$, analytic in the interior, such that
\[
w_2^\omega(F, \Psi_{t+i\beta}(F')) = w_2^\omega(\Psi_t(F'), F) \quad \quad (t \in \mathbb{R})\,.
\]

\begin{definition}\label{def:musc}
We say that a state $\omega$ on $\mathcal{A}$ is {\it admissible at} $x\in M$ if
\[
\WF(\boldsymbol{\phi}(.)\Omega)\cap T_x^*M \supset \mathcal{N}^-_x,
\]
where $\mathcal{N}_x^-=\mathcal{N}^-\cap T_x^*M$ and
$\mathcal{N}^-\subset T^*M$ is the set of past-directed lightlike co-vectors. We call
$\omega$ {\it admissible} if it is admissible at all $x\in M$, or equivalently when
\[
\WF(w_2^\omega) \supset \left\{(x,-k;x,k) \ |\  (x,k)\in\mathcal{N}^- \right\}.
\]
\end{definition}
We note that $\WF(\boldsymbol{\phi}(.)\Omega)$ is the wave front set of the Hilbert space-valued 
distribution $F\mapsto \boldsymbol{\phi}(F)\Omega$, cf.\ \cite{SVW}. The equivalence in the definition
follows from Prop.\ 6.1 of \cite{SVW}. As discussed in the Introduction, for the quantized Klein-Gordon field the
microlocal spectrum condition at the level of 2-point functions  implies this admissibility condition
(cf.\ eqn.\ \eqref{eqn:WF2}) and 
this holds also for other types of quantized fields that are subject to a linear hyperbolic equation, like the Dirac
field and the vector potential \cite{SV2}.

\bigskip

Our admissibility condition can even be expected for many reasonable interacting theories.
Indeed, let us suppose that the theory is asymptotically free in the sense that at every point $x\in M$
the state $\omega$ has a short-distance scaling limit $\omega_x$ at $x$ \cite{FH}, such that the scaling limit state
$\omega_x$ is the vacuum state of a free quantized field on $T_xM \simeq$ Minkowski spacetime. 
Combining Prop.\ A.2 of \cite{Sa2} (see also Prop.\ 2.8 of \cite{SV2}, cf.\ also \cite{VerACS}) for scaling limits 
of distributions with the results of \cite{SVW} we find in this case
\[
\WF(\boldsymbol{\phi}(.)\Omega) \supset \{(x,k) \;|\; (0,k)\in \WF(\boldsymbol{\phi}_0(.)\Omega_x)\}
\supset \mathcal{N}^-\cap T_x^*M.
\]
where $\boldsymbol{\phi}_0$ refers to the scaling limit quantum field.
Taking the union over all $x\in M$ we conclude that $\omega$ is admissible.

\bigskip

\begin{theorem*}
Consider a KMS state $\omega$ at inverse temperature $\beta$ with respect to the evolution induced by the complete smooth vector field $V$
on $E$ with projection $\chi$ on $M$. If $\chi$ is spacelike at a point $x\in M$, then $\omega$ is not admissible at the point $x$.
\end{theorem*}
\noindent
For quantum fields satisfying a linear hyperbolic equation, the proof of this statement is actually contained in the proof
of Thm.\ 5.1 of \cite{SV} (the assumption made in this reference that the Killing vector field is timelike does not enter in
the proof). However, here we give a different proof.
\begin{proof}
Since every $\beta$-KMS state is invariant under the evolution $\Psi_t$, there is a strongly continuous unitary group 
$U(t)=e^{itH}$ on $\mathcal{H}$, generated by a self-adjoint operator $H$, such that
\[
e^{itH}\boldsymbol{\phi}(F)\Omega= \boldsymbol{\phi}(\Psi_t(F))\Omega
\]
for all $F\in\mathcal{D}(E)$. Moreover, due to the $\beta$-KMS condition, the map
\[
t\mapsto e^{itH}\boldsymbol{\phi}(F)\Omega
\]
is bounded and continuous on the strip $S_{\beta/2}$ and holomorphic on the interior (\cite{Ped} Prop.~8.14.2).

Now consider a coordinate system $(\xi,\mathcal{O})$ which contains $x$ and is adapted to $\tau_t$. 
This means that for any point $y\in\mathcal{O}$ with $\xi(y)=(y^0,y^1,y^2,y^3)$ we have
\[
\xi(\tau_t{y}) = (y^0,y^1+t,y^2,y^3), \qquad \text{ for all } t\in \mathbb{R} \text{ such that } \tau_t y\in  \mathcal{O}.
\]
Shrinking $\mathcal{O}$ if necessary we may also introduce a frame $\{e_1,\ldots,e_N\}$
for $E$ on $\mathcal{O}$ such that $\Psi_t(e_j)=e_j$ for each $j=1,\ldots,N$ at all points where both sides are defined.
(We may first choose such a frame on the hypersurface $y^1=0$ and then use the evolution $\Psi_t$ to extend it.)
For each $j=1,\ldots,N$ we consider the $\mathcal{H}$-valued distribution $G_j$ on $(0,\beta/2)\times \mathcal{O}$ defined by
\[
G_j(\eta,f)=e^{-\eta H}\boldsymbol{\phi}(fe_j)\Omega,
\]
where $f$ is in $\mathcal{D}(\mathcal{O})$. $G_j$ is seen to satisfy $(\partial_y+i\partial_{\eta})G_j=0$,
so it is holomorphic in $y+i\eta$, and its boundary value as $\eta\to 0$ is the distribution $f\mapsto \boldsymbol{\phi}(fe_j)\Omega$.
Although $G_j$ is not a smooth function, we can still conclude as in Thm.\ 2.8 of \cite{SVW} that
\[
\WF(\boldsymbol{\phi}(.e_j)\Omega)\cap T_x^*M\subset \{(x,k)\in T^*M \;|\; k(\chi)\ge 0\}\,;
\]
in fact, the proof of  Thm.\ 2.8 of \cite{SVW} can be seen by inspection to generalize to the case that the $G_j(\,.\,)$ are distributions
analytic in $y + i\eta$, on observing that $|G_j(\eta,y,f)| \le C||f||_m$ for a Sobolev norm of sufficiently high degree $m$,
for all $f(y^0,y^2,y^3)$ supported in a fixed compact set and a constant $C > 0$, uniformly in $(\eta,y)$. (See also
Thm.\ 8.1.6 of \cite{Hormander}.)

Combining the components $\boldsymbol{\phi}(.e_j)\Omega$ of the distribution $\boldsymbol{\phi}(.)\Omega$ we find (cf.\ \cite{SV2})
\[
\WF(\boldsymbol{\phi}(.)\Omega)\cap T_x^*M\subset \bigcup_{j=1}^N \WF(\boldsymbol{\phi}(.e_j)\Omega)\cap T_x^*M
\subset \{(x,k)\in T^*M \;|\; k(\chi)\ge 0\}.
\]
However, when $\chi$ is spacelike at $x$, there is a dual
null vector $k$ at $x$ such that $k(\chi)<0$, so we cannot have $\mathcal{N}^-\subset \WF(\boldsymbol{\phi}(.)\Omega)$, 
i.e.\ $\omega$ cannot be admissible at $x$.
\end{proof}

\section{The case of Fermi fields on flat spacetime} \label{fermifields}

In this section we shall discuss the existence of KMS states for Fermi fields with respect to spacelike translations in a four dimensional Minkowski spacetime. The state we shall obtain will however be inadmissible in the sense of Section \ref{Sec:Result}.

In order to construct the desired KMS state for free Fermi fields we need just consider the two-point function, because we choose the  state to be quasi-free.
This two-point function can be explicitly constructed as follows. Consider a free Dirac field $\psi$ on Minkowski spacetime.\footnote{We refer to \cite{Sa2} for details about the quantization of Dirac fields and for the form of the distribution $S$ implementing the CAR.}
The two-point function of a KMS state $\omega$ with respect to spacelike translations along the direction determined by the 
normalised spatial vector $e_j$ and implemented by the one parameter group of $*-$automorphisms $\alpha_s(\psi(x))= \psi(x-se_j)$ can be constructed out of the anticommutator function $S(x,y)=\omega(\psi(x)\psi^\dagger (y))+\omega(\psi^\dagger{(y)}\psi(x))$ which defines the CAR relations.
Since $S$ is a translation-invariant Schwartz distribution we can consider its Fourier transform $\hat{S}$ and multipliying it with the corresponding Fermi factors we obtain  
\[
{{\hat{w}}_2^{\omega +}}(k) = \frac{\hat{S}(k)}{e^{-\beta k_j}+1 }\;, \qquad
{{\hat{w}}_2^{\omega -}}(k) = \frac{\hat{S}(k)}{e^{+\beta k_j}+1 }\;,
\]
where $k_j=\langle k,e_j\rangle$. The functions $e^{\pm iz k_j}{(e^{\mp\beta k_j}+1 )}^{-1}$ are smooth, bounded functions on $\mathbb{{R}}^4$ for every 
$\text{Im}(z)\in[0,\beta]$, furthermore, they are positive for $z=0$, hence also  $\hat{w}_2^{\omega \pm}(k)$ are Schwartz distributions. Furthermore, their inverse Fourier transforms 
define the two-point functions of a quasi-free state
$\omega(\psi(x)\psi^\dagger(y))={w_2^{\omega +}}(x-y) $ and $\omega(\psi^\dagger(y)\psi(x))={w_2^{\omega -}}(x-y) $.
This state is a KMS state with respect translations along the direction $e_j$ at inverse temperature $\beta$, however it cannot be a Hadamard state as proved in Section \ref{Sec:Result}.

For quantized bosonic fields on Minkowski spacetime fulfilling a linear hyperbolic field equation, a similar procedure cannot be applied.
Actually, if one tries to construct the relevant two-point functions of extremal KMS states multiplying the commutator function with the appropriate Bose factor in the Fourier domain along the lines of \cite{BV-unruh1}, some divergences are encountered.
These divergences are due to the fact that the Bose factor with respect to spacelike translations is not locally integrable in momentum space.

\section{Summary and Outlook}
We have shown that for quantum fields on curved spacetimes obeying minimal assumptions, a microlocal admissibility condition for the 
2-point function of a state is locally incompatible with the KMS condition with respect to a {\it spacelike} evolution group.
For free or asymptotically free fields this implies an incompatibility with the microlocal spectrum condition, which is stronger.

As indicated, there are examples of spacetimes that are considered to be of physical relevance, such as the rotating Kerr black hole spacetimes,
possessing Killing vector fields which are timelike in some  (accessible) region of spacetime and spacelike in other (accessible) regions. Therefore, quantum field states that fulfill the admissibility condition can only satisfy the KMS condition in the regions of spacetime where the Killing vector field of 
the evolution group is not spacelike. There are examples for such behaviour already in Minkowski spacetime: By the Bisognano-Wichmann Theorem \cite{BiWi},
the vacuum state of any Wightman-type quantum field theory restricts to a KMS state at inverse temperature $2\pi$ with respect to the action of a one parametric group 
of Lorentz boosts on the algebra of field operators localized in the Rindler wedge region $W$ of spacetime containing all boost trajectories such that the associated Killing vector field is timelike and 
future-directed. On the algebra of the causal complement $-W$ of that region, the vacuum state restricts to a KMS-state at inverse temperature $-2\pi$ with respect to
the same one-parametric group of Lorentz boosts. In the (open) complement of $W \cup -W$ in Minkowski spacetime, the trajectories of the one-parametric group
of Lorentz boosts are spacelike, and the vacuum state has no KMS-like properties in restriction to operators localized in that complement region. A completely analogous situation occurs for quantum fields on the Schwarzschild-Kruskal spacetime \cite{Ka85,KW,Sa15}.
In the case of the Kerr rotating black hole spacetime,
where the Killing vector field in the exterior of the black hole region changes between timelike and spacelike, it is still possible
that there exist quantum field states that fulfill the admissibility and microlocal spectrum condition and have KMS-like properties in the region where the Killing vector field is timelike (a local generalization of the KMS condition has been given, see e.g.\ \cite{GPV} and references cited there).
Our method of proof does a priori not rule out the possibility of 
quantum field states that fulfill the admissibility and microlocal spectrum condition and the KMS condition with respect to an evolution group with a lightlike Killing vector field, and it would be interesting to investigate that possibility further.

\subsection*{Acknowledgements}
N.P.\ thanks the Institute for Theoretical Physics of the University of Leipzig for the 
kind hospitality during the preparation of this work and the DAAD for supporting this visit with the program ``Research Stays for Academics 2017''.
K.S. thanks Christian G\'erard for bringing his notes on this result to the attention of N.P. and R.V.

\end{document}